\documentclass[twocolumn,showpacs,preprintnumbers,amsmath,amssymb,floatfix]{revtex4}

\usepackage{graphicx}
\usepackage{dcolumn}
\usepackage{bm}

\newcommand{\beq}{\begin{equation}}
\newcommand{\eeq}{\end{equation}}
\newcommand{\bey}{\begin{eqnarray}}
\newcommand{\eey}{\end{eqnarray}}

\begin{document}

\title{BTZ black holes inspired by noncommutative geometry}

\author{Farook Rahaman}
\email{rahaman@iucaa.ernet.in} \affiliation{Department of
Mathematics, Jadavpur University, Kolkata 700 032, West Bengal,
India}

\author{P.K.F. Kuhfittig}
\email{kuhfitti@msoe.edu} \affiliation{Department of Mathematics,
Milwaukee School of Engineering, Milwaukee, Wisconsin 53202-3109,
USA}

\author{B.C. Bhui}
\email{bikas_bhui@rediffmail.com} \affiliation{Department of
Mathematics, Meghnad Saha Institute of Technology, Kolkata-700150,
India}

\author{Masiur Rahaman}
\email{mosiurju@gmail.com} \affiliation{Department of Mathematics,
Meghnad Saha Institute of Technology, Kolkata-700150, India}

\author{Saibal Ray}
\email{saibal@iucaa.ernet.in} \affiliation{Department of Physics,
Government College of Engineering and Ceramic Technology, Kolkata
700 010, West Bengal, India}

\author{U.F. Mondal}
\email{umarfarooquemondal@ymail.com} \affiliation{Department of
Mathematics, Behala College,  Parnasree, Kolkata 700060, India}

\date{\today}

\begin{abstract}\noindent
In this paper a Ba$\tilde{\text{n}}$ados, Teitelboim and Zanelli
(BTZ) \cite{BTZ92} black hole is constructed from an exact
solution of the Einstein field equations in a (2+1)-dimensional
anti-de Sitter spacetime in the context of noncommutative
geometry.  The BTZ black hole turns out to have two horizons, no
horizon or a single horizon corresponding to a minimal mass.
Certain thermodynamical properties are investigated, including
Hawking temperature, entropy and heat capacity.  Also discussed is
the geodesic structure of BTZ black holes for both massless and
massive particles.  In particular, it is shown that bound orbits
for test particles are possible.
\end{abstract}


\maketitle

\section{Introduction}\noindent
Recent years have seen rapid advances in the applications
of noncommutative geometry, an outgrowth of string
theory.  The approach is based on the realization that
coordinates may become noncommuting operators on a $D$-brane
\cite{SS03,NSS06,NS10,eS08}.  The result is a discretization
of spacetime due to the commutator
$[\textbf{x}^{\mu},\textbf{x}^{\nu}]=i\theta^{\mu\nu}$, where
$\theta^{\mu\nu}$ is an antisymmetric matrix.  It is similar
to the way that the Planck constant $\hbar$ discretizes phase
space \cite{SS03}.  The noncommutativity eliminates point-like
structures and replaces them with smeared objects.  The
noncommutative geometry is an intrinsic property of spacetime
and does not depend on particular features such as curvature.

A number of studies inspired by noncommutative geometry can
be found in the literature.  In one of the earlier studies,
Garattini and Lobo \cite{GL09} obtained exact wormhole
solutions that were then analyzed in semi-classical gravity.
In a subsequent study \cite{LG10} they found an exact
gravastar solution and explored its physical characteristics.
Rahaman et al. \cite{RKCUR12}, discussing galactic rotation
curves, concluded that a noncommutative-geometry background
is sufficient for producing stable circular orbits without
the need for dark matter.  Kuhfittig \cite {pK12} found
that a special class of thin-shell wormholes that are
unstable in classical general relativity exhibit small
regions of stability in noncommutative geometry.  In another
study by Rahaman et al. \cite{frprd} on higher-dimensional
wormholes, it is shown that wormhole solutions exist in the
usual four, as well as in five dimensions, but they do not
exist in higher-dimensional spacetimes.  In a more recent
study, Radinschi et al. \cite{irina} calculated the
energy-momentum for a noncommuting radiating Schwarzschild
black hole in order to obtain the expression for energy.
Common to all these studies is that the effect of the
smearing is mathematically implemented by using a Gaussian
distribution of minimal length $\sqrt{\theta}$ instead of
the Dirac delta function.

Interest in $(2 + 1)$-dimensional gravity has increased in recent
years due to the discovery of various aspects of black-hole
solutions. Some general works in this line are: quasinormal modes
of charged dilaton black holes in $(2 + 1)$-dimensional solutions
in low-energy string theory with asymptotically anti-de Sitter
spacetimes \cite{Fernando}, branes with naked singularities,
analogous to linear or planar defects in crystals and showing that
zero-branes in AdS spacetimes are ``negative mass black holes''
\cite{Zanelli}, Hawking radiation from covariant anomalies in $(2
+ 1)$-dimensional black holes \cite{Nam} and so on. On the other
hand, specialized investigations have been carried out by Rahaman
et al. \cite{Rahaman} on gravastars in $(2 + 1)$ anti-de Sitter
spacetimes with charge as an alternative to charged black holes.
Also, nonstatic charged BTZ-like black holes in $(N +
1)$-dimensions have been considered by Ghosh \cite{Ghosh}, which
in the static limit, for $N = 2$, reduces to $(2 + 1)$ BTZ black
hole solutions. There are also charged, regular extensions of the
BTZ black hole solutions available in the literature by employing
nonlinear Born-Infeld electrodynamics to eliminate the inner
singularity \cite{Mazharimousavi}.

  There are several proposals  in the literature for
constructing noncommutative black holes \cite{LO,OM,MR,MY}.
However, following Nicolini et al \cite{NSS06},  we construct a
BTZ black-hole solution from the Einstein field equations in (2 +
1)-dimensional anti-de Sitter spacetime, given a
noncommutative-geometry background. This is followed by a
discussion of the black hole's thermodynamical properties such as
Hawking temperature, entropy and heat capacity, as well as the
geodesic structure for both massless and massive particles.  Some
of the thermodynamical properties are similar to those obtained by
Liang and Liu \cite{jL12} who used a Lorentzian smeared mass
distribution instead of a Gaussian one in $AdS_3$ spacetime.  One
consequence of this is that in the limit,
$r/\sqrt{\theta}\rightarrow \infty$, the solution reduces to a
rotating BTZ black hole. Returning to a Gaussian framework,
rotating black holes in (2+1) dimensions are also discussed in
Ref. \cite{TL12} by Tejeiro and Larranaga. They \cite{LT11} also
describe charged black holes and compare them to charged BTZ black
holes.

\section{The interior spacetime}\noindent
Let us write the line element describing  the interior space-time
of a static spherically symmetric distribution of matter in
$(2+1)$ dimensions in the following form:
\begin{equation}\label{E:line1}
ds^2 = -f(r) dt^2 + [f(r)]^{-1} dr^2 + r^2d\phi^2,
\end{equation}
where $f(r)$ is denoted by $e^{2\nu(r)}$ and $[f(r)]^{-1}$
by $e^{2\mu(r)}$.  We take the matter distribution to be
anisotropic in nature and therefore choose the most general
energy-momentum tensor in the form
\begin{equation}\label{E:em}
T_{ij} = (\rho + p_t) u_i u_j + p_t g_{ij} + (p_r - p_t)\chi_i
\chi_j,
\end{equation}
where $\rho$, $p_r$ and $p_t$ represent the energy density, radial
pressure and tangential pressure, respectively.  Also, $\chi^{i} =
e^{-\mu(r)}\delta^i_r$ is a unit four vector along the radial
direction and  $u^{i}$ is the $4$-velocity of the fluid.

The Einstein field equations with cosmological constant ($\Lambda
< 0$), together with the general energy-momentum tensor given in
Eq.~(\ref{E:em}), yield (letting $G = c = 1$)

\begin{eqnarray}
2\pi \rho +\Lambda &=& \frac{\mu' e^{-2\mu}}{r}, \label{eq3} \\
2\pi p_r -\Lambda &=& \frac{\nu' e^{-2\mu}}{r}, \label{eq4}  \\
2\pi p_t -\Lambda &=&
e^{-2\mu}\left(\nu'^2+\nu''-\nu'\mu'\right).\label{eq5}
\end{eqnarray}
We have, in addition, the conservation equation in $(2+1)$
dimensions:
\begin{equation}\label{E:conservation}
\left(\rho + p_r\right)\nu' + p_r' + \frac{1}{r}\left(p_r -
p_t\right) =0.
\end{equation}

In $(2+1)$ dimensions, the maximally localized source of energy of
the static and spherically symmetric distributions having a
minimal spread Gaussian profile is taken as \cite{eS08}
\begin{equation}\label{E:density1}
\rho = \frac{M}{4 \pi \theta}~\text{exp}
\left(-\frac{r^2}{4\theta}\right).
\end{equation}
Here $M$ is the total mass of the source. Due to the uncertainty,
it is diffused throughout a region of linear dimension
$\sqrt{\theta}$.

The vacuum Einstein field equations in $(2+1)$ spacetime
dimensions, with a negative cosmological constant, admit a black
hole solution known as a BTZ solution \cite{BTZ92}. For a  BTZ
black hole, we have $g_{rr} =g_{tt}^{-1}$.  So to retain the
structure, we require that
\begin{equation}
p_r = -\rho.
\end{equation}
This {\it ansatz} is known in the literature as a `$\rho$-vacuum'
or `vacuum equation of state' in connection with the `zero point
energy of quantum fluctuation' \cite{Blome,Davies,Hogan,Kaiser}
where pressure is of a repulsive nature.

With this equation, one can solve Eq. (\ref{E:conservation})
to yield
\begin{equation}\label{E:pt}
p_t =\frac{M\left( \frac{r^2}{2 \theta} -1\right )}{4 \pi
\theta}~\text{exp} \left(-\frac{r^2}{4\theta}\right).
\end{equation}
Using the field equations, we get the following solution
for the metric coefficients:
\begin{equation}\label{E:gtt}
e^{-2\mu} = e^{2\nu} = -A+ 2M~\text{exp}
\left(-\frac{r^2}{4\theta}\right) - \Lambda r^2,
\end{equation}
where $A$ is an integration constant.

In the limit, $\frac{r}{\sqrt{\theta}}  \rightarrow \infty$,
so that Eq. (\ref{E:gtt}) reduces to a BTZ black hole, where
the integration constant $A$ plays the role of the mass of
the BTZ black hole, i.e., $A=M$. Observe that asymptotically
far away, $\rho = p_r =p_t =0$.

To determine the mass distribution from  Eq.
(\ref{E:density1}), we use an approach similar to that in
Refs. \cite{eS08} and \cite{GL09}:
\begin{equation}\label{E:mass1}
 m(r)=\frac{M}{\pi^{({\tilde m} - 2)/2}}\gamma
  \left[\frac{\tilde m}{2},\alpha^2\left(\frac{r}
  {2M}\right)^2\right],
\end{equation}
where $\alpha^2=M^2/\theta$ and $\gamma$ is the lower
incomplete gamma function
\begin{equation}\label{E:gamma}
     \gamma\left(\frac{a}{b},x\right)=
     \int^x_0u^{a/b}e^{-u}\frac{du}{u}.
\end{equation}\label{E:mass2}

For a BTZ black hole, ${\tilde m} = 2$, and we obtain from
$\alpha^2 r^2/4M^2=r^2/4\theta$ the expression for mass as
\begin{equation}\label{E:mass2}
  m(r)=M\int^{r^2/4\theta}_0e^{-t}dt=
  M \left[ 1 -~\text{exp}
\left(-\frac{r^2}{4\theta}\right)\right].
\end{equation}
The parameter $\alpha$ plays a critical role in determining the
horizons, as we will see later on.  At the origin, $m(0)=0$, which
is consistent with Eq. (\ref{E:mass1}).

Near the origin, the geometry is given by
\begin{equation*}
e^{-2\mu} = e^{2\nu} = -A +2M - \left(\frac{2M  }{4 \theta}
+\Lambda \right) r^2 +\mathcal{O}(r^4).
\end{equation*}
One can identify this result with a BTZ black hole spacetime,
where the total mass $M$ and the noncommutative parameter
$\theta$ combine to modify the cosmological constant, a point
also made in Ref. \cite{eS08}.  This indicates that different
mass particles experience different cosmological constants.
We therefore conclude that our line element describes the
geometry of a noncommutative-geometry inspired BTZ black
hole.

\section{Features of the black hole}\noindent
In this section we study some of the effects of the
noncommutative geometry on BTZ black holes.  Let $A=M$ in Eq.
(\ref{E:gtt}).  Then the equation $g_{tt}(r_h) =0$ gives the
event horizon(s):
\begin{equation}\label{Rad1}
r_h^2 = \frac{M}{\Lambda}\left[2~\text{exp}
\left(-\frac{r_h^2}{4\theta}\right) - 1\right].
\end{equation}
Even though we cannot obtain a closed-form solution for $r_h$ in
Eq. (\ref{Rad1}), we can readily write the mass $M$ as a function
of $r_h$:
\begin{equation}
M = \frac{\Lambda r_h^2}{2\text{exp}
\left(-\frac{r_h^2}{4\theta}\right) -1}.
\end{equation}

The existence of horizons and their radii can be found at the
points where $g_{tt}$ cuts the $r$-axis, as shown in Fig. 1, using
$\Lambda\sqrt{\theta}=-0.02$.  Here three possibilities present
themselves graphically in terms of the approximate value of
$\alpha$:\\ (i) two horizons when $\alpha>0.214$, or  $M> M_0 =
0.214\sqrt{\theta} $;\\ (ii) one horizon corresponding to the
extremal black hole with $M= M_0$,
  i.e., $\alpha=0.214$;\\
(iii) no horizon for $\alpha<0.214$.\\
Fig. 1 shows that a noncommutative-geometry inspired BTZ black
hole has two horizons and that the distance between the horizons will
increase with an increasing  black-hole mass. Fig. 1 also indicates
that there is  a minimal mass $M_0$ below which no black hole exists.
Moreover, at the minimal mass  $M=M_0$, the two horizons coincide at
the minimal horizon radius $r_0$, which  lies between the horizons.
This $r_0$ is therefore the horizon radius of the extremal black hole.
It can also be determined from the conditions $f=0$ and
$\frac{df}{dr}=0$, leading to the equation
\begin{equation}
  2\,\text{exp}
\left(\frac{r_0^2}{4\theta}\right) +\frac{r_0^2/\theta}{1-2 \text{exp}
\left(-\frac{r_0^2}{4\theta}\right)}=0.
\end{equation}
Using the condition $\Lambda\sqrt{\theta}=-0.02$ in Fig. 1, we
can obtain $r_0/\sqrt{\theta}=2.59$, showing that the two approaches
are consistent.

The minimal mass of the extremal black hole can be written in
terms of the minimal radius $r_0$, so that

\begin{equation}
  \frac{M_0}{\sqrt{\theta}}= \frac{\frac{r_0^2}{\theta}
  ( \Lambda\sqrt{\theta})}{2\,\text{exp}
  \left(-\frac{r_0^2}{4\theta}\right)-1}=0.214.
\end{equation}
The variation of this factor with respect to the horizon radius is
shown in Fig. 2. At this point let us also plot
$\rho\sqrt{\theta}$ from Eq. (\ref{E:density1}) for various values
of $M/\sqrt{\theta}$ (Fig. 3).  Similarly, Fig. 4 shows
$p_t\sqrt{\theta}$ plotted against $r/\sqrt{\theta}$ from Eq.
(\ref{E:pt}).

It is worth noting that the radius of the  extremal black hole
is always less than the radius of the outer horizon. The
result significant: if the initial mass of the black hole is
$M>M_0$, then it can radiate until the value $M_0$ is reached.
It follows that evaporation of the black hole may indeed
be occurring.

 \begin{figure}
\includegraphics[scale=.3]{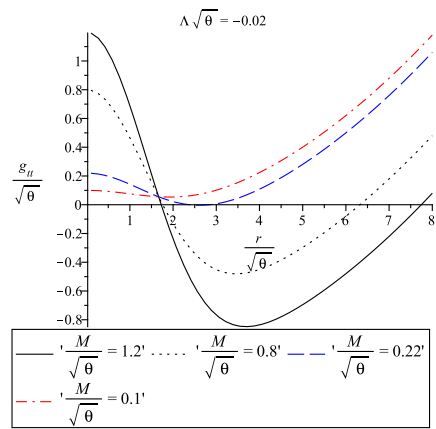}
\caption{The singularities occur where  $g_{tt}$ cuts $r$-axis.
Representation: the
solid curve for $M=1.2 \sqrt{\theta}$ and dotted curve for $M=0.8
\sqrt{\theta}$   indicate two horizons. The dashed curve for $M=0.214
\sqrt{\theta}$ represents one degenerate horizon $r_0
\approx 2.59\sqrt{\theta} $, i.e., an extremal black hole. For $M=0.1
\sqrt{\theta}$, no horizon exists (chain curve). The intercepts
on the $r$-axis give the radii of the event horizons.}
\end{figure}
\begin{figure}
\includegraphics[scale=.3]{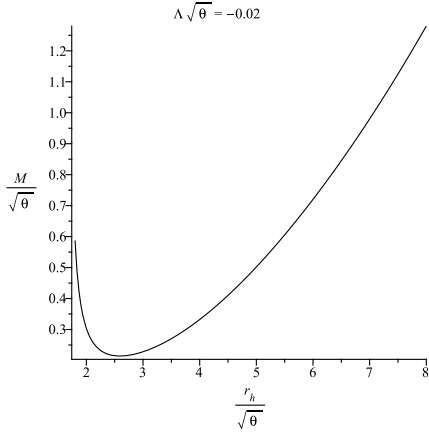}
\caption{The variation of  mass $\frac{M}{\sqrt{\theta}}$ with
respect to horizon radius $\frac{r_h}{\sqrt{\theta}}$.  }
\end{figure}

\begin{figure}
\includegraphics[scale=.3]{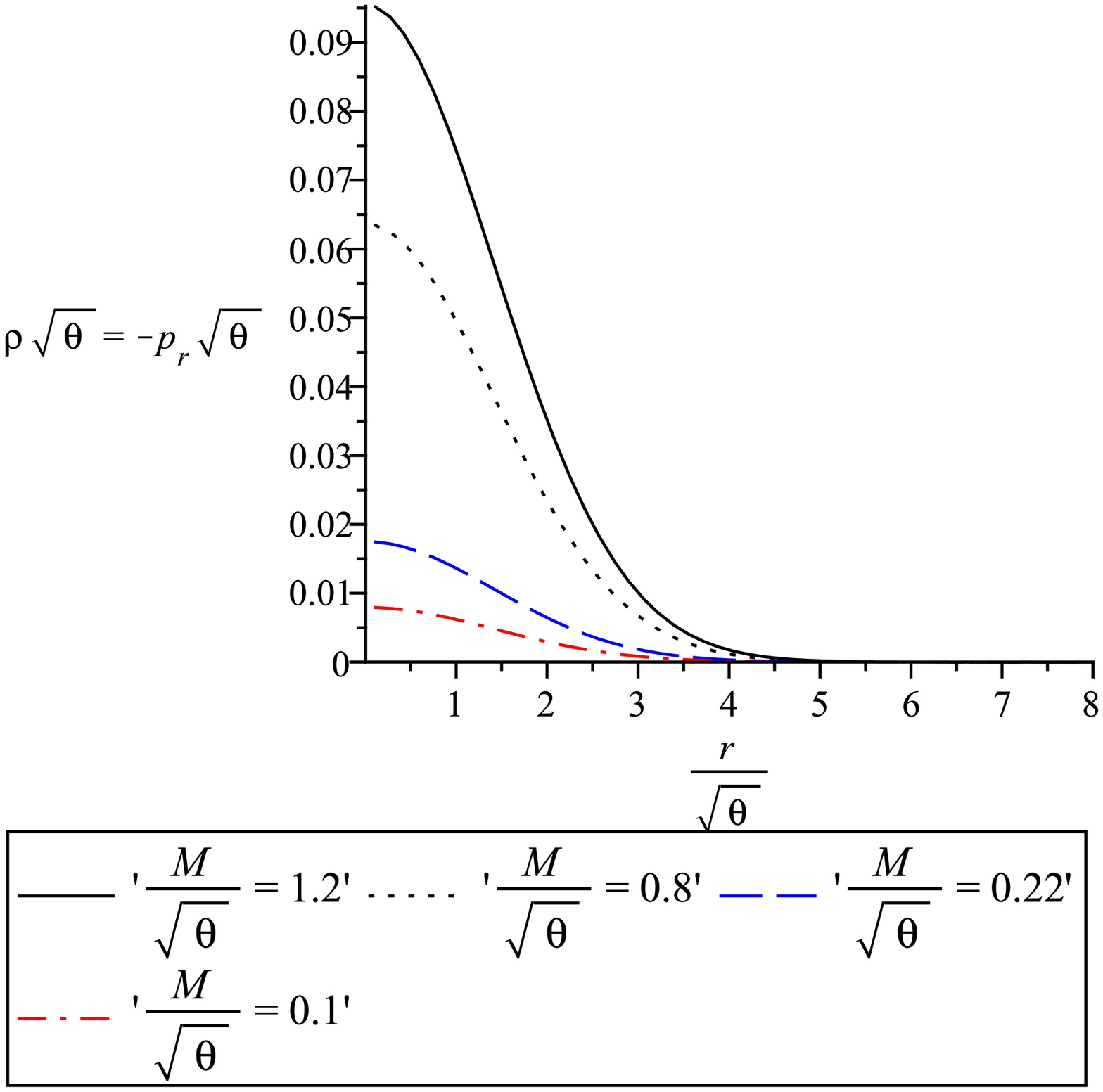}
\caption{Plot for $\rho \sqrt{\theta}$ vs.
$\frac{r}{\sqrt{\theta}}$ for different values of
$\frac{M}{\sqrt{\theta}}$.  Representation: the solid curve for
$M=1.2\sqrt{\theta}$,  dotted curve for $M=0.8 \sqrt{\theta}$,
dashed curve for $M=0.214 \sqrt{\theta}$, and chain curve
for $M=0.1 \sqrt{\theta}$.}
\end{figure}

\begin{figure}
\includegraphics[scale=.3]{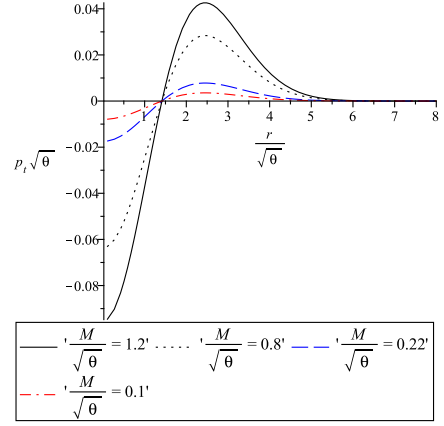}
\caption{Plot for $p_t \sqrt{\theta}$ vs.
$\frac{r}{\sqrt{\theta}}$ for different values of
$\frac{M}{\sqrt{\theta}}$. Representation: the solid curve for
$M=1.2\sqrt{\theta}$,  dotted curve for $M=0.8 \sqrt{\theta}$,
dashed curve for $M=0.214 \sqrt{\theta}$, and chain curve for $M=0.1
\sqrt{\theta}$. Note that up to the degenerate horizon $r_0
\approx 2.59\sqrt{\theta} $ for an extremal black hole, $p_t$ assumes
negative values and beyond that positive values. For
large values of $r$, i.e., for $\frac{r}{\sqrt{\theta}} > > 1$,
   it dies out.}
\end{figure}

Next, consider the Hawking temperature, which is given by
\begin{multline}\label{E:Hawking1}
T_H = \frac{1}{4 \pi} \left(\frac{d g_{tt}}{dr} \right)
\sqrt{-g^{tt}g^{rr}}|_{r=r_h} \\= -\frac{r_h}{4 \pi} \left[2 \Lambda
+\frac{M}{\theta}~\text{exp}
\left(-\frac{r_h^2}{4\theta}\right)\right].
\end{multline}
The plot, Fig. 5, is obtained from
\begin{equation}\label{E:Hawking2}
T_H = -\frac{r_h/\sqrt{\theta}}{4 \pi} \left[2 \Lambda
\sqrt{\theta}+\frac{M}{\sqrt{\theta}}~\text{exp}
\left(-\frac{r_h^2}{4\theta}\right)\right].
\end{equation}

Eq. (\ref{E:Hawking2}) and Fig. 5 show that the noncommutative
geometry leads to the minimal horizon radius $r_0$, since $T_H$
cannot be negative.  This is exactly where $\frac{dg_{tt}}{dr}=0$
in Eq. (\ref{E:gtt}). Observe that the temperature sinks to
absolute zero at $r_0$. For $\frac{r_h}{\sqrt{\theta}} >
> 1$, the Hawking temperature assumes the value
\begin{equation}
T_H \approx  -\frac{r_h \Lambda}{2 \pi}.
\end{equation}

\begin{figure}
\includegraphics[scale=.3]{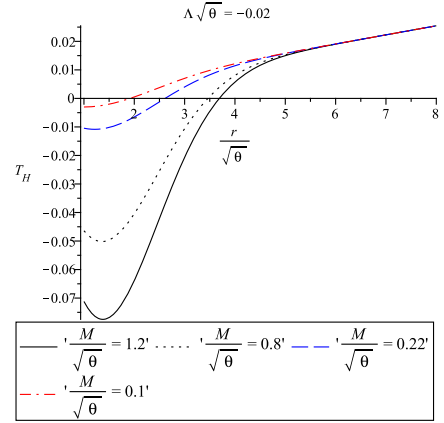}
\caption{Plot for Hawking temperature vs.
$\frac{r}{\sqrt{\theta}}$ for different values of
$\frac{M}{\sqrt{\theta}}$. Representation: the solid curve for $M=1.2
\sqrt{\theta}$,  dotted curve for $M=0.8 \sqrt{\theta}$, dashed curve
for $M=0.214 \sqrt{\theta}$, and chain curve for $M=0.1
\sqrt{\theta}$. }
\end{figure}

For completeness, let us also state the closely related
surface gravity
\begin{equation}
\kappa = \frac{1}{2} \left(\frac{d g_{tt}}{dr} \right)
 |_{r=r_h} = -\frac{r_h}{2} \left[2 \Lambda
+\frac{M}{\theta}~\text{exp}
\left(-\frac{r_h^2}{4\theta}\right)\right],
\end{equation}

\begin{figure}
\includegraphics[scale=.3]{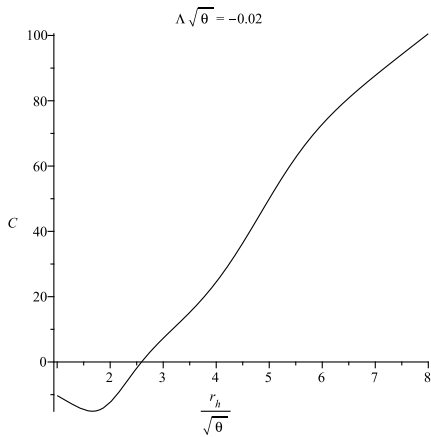}
\caption{Plot for C vs. $\frac{r_h}{\sqrt{\theta}}$.}
\end{figure}as well as the Bekenstein-Hawking entropy (S) of
the black hole.  It is twice the perimeter $L$ of the event
horizon:
\begin{equation}
S = 2 L = 4 \pi r_h.
\end{equation}

As a final comment, the noncommutative geometry inspired BTZ
black hole is stable if the heat capacity $C$ is positive
\cite{jL12}, where
\[
   C= \frac{\partial M(r_h)}{\partial T(r_h)}=
   \frac{\partial M(r_h)}{\partial r_h}\frac{1}
   {\frac{\partial T(r_h)}{\partial r_h}}.
\]

The nature of the heat capacity is shown in Fig. 6. The plot shows
that $C$ vanishes at the extremal event horizon $r_0$ and becomes
negative for $\frac{r_h}{\sqrt{\theta}} < \frac{r_0}{\sqrt{\theta}} $,
just as in the case of the Hawking temperature.  So this region is
definitely unphysical.  On the other hand, for
$\frac{r_h}{\sqrt{\theta}}>\frac{r_0}{\sqrt{\theta}} $,
$C$ is positive, which implies that the BTZ black hole is
stable.

 At this point we would like to comment on  how the
physical quantities such as temperature, entropy, etc., could be
affected by the noncommutativity
 for small $\theta$. We calculate the different physical quantities   of the standard BTZ
black hole plus perturbative terms. The BTZ black hole has the
horizon located at
\begin{equation} r_h = \sqrt{M} l , \end{equation} where $ \Lambda =
-\frac{1}{l^2}$ to emphasize that $\Lambda$ is negative.

Equation (14) can be solved by iteration. The result is
\begin{equation} r_h = \sqrt{M} l \left[ 1- 2~ exp
\left(-\frac{Ml^2}{4\theta}\right)\right]^{\frac{1}{2}}.
\end{equation}
For small $\theta$ and   $\frac{r_h}{\sqrt{\theta}} > > 1$, the
above equation can be written as
\begin{equation} r_h \approx \sqrt{M} l \left[ 1- ~ exp
\left(-\frac{Ml^2}{4\theta}\right)\right].
\end{equation}
 Here the first term  is the BTZ black hole    horizon radius,
while the second term is the $\theta$ correction.

It now becomes apparent that the effect of noncommutativity is
small, as expected,
 because
spacetime should be  a smooth classical manifold at large
distances.

The next step is to find the  $\theta$ corrections of the
Bekenstein-Hawking entropy (S) of the BTZ black hole:
\begin{equation} S = 4 \pi r_h \approx 4 \pi \sqrt{M} l  - 4 \pi \sqrt{M} l ~ exp
\left(-\frac{Ml^2}{4\theta}\right) .\end{equation} Here the first
term  is the Bekenstein-Hawking entropy (S) of the BTZ black hole,
while the second term is the $\theta$ correction.

Our final task is to find the $\theta$ corrections for the
Hawking temperature and surface gravity of the BTZ black hole.

From eq. (18), we have,
\begin{widetext}
\begin{eqnarray}
T_H = -\frac{1}{4 \pi } \left[ \sqrt{M} l \left\{ 1- ~ exp
\left(-\frac{Ml^2}{4\theta}\right)\right \} \right]
  \left [-\frac{2}{l^2} +\frac{M}{\theta} ~exp
\left\{-\frac{M l^2}{4 \theta }
 \left[ 1- ~2 exp \left(-\frac{Ml^2}{4\theta}\right)\right]
\right\} \right] .  \end{eqnarray}
\end{widetext}
Keeping the first order of $exp
\left(-\frac{Ml^2}{4\theta}\right)$, the above expression yields
\begin{widetext}
\begin{eqnarray}T_H \approx \frac{\sqrt{M}}{2 \pi l } - \left[ \frac{M \sqrt{M} l}{4 \pi \theta } exp \left\{-\frac{M l^2}{4
\theta } \right\} + \frac{ \sqrt{M}  }{ 2 \pi l  } exp
\left\{-\frac{M l^2}{4 \theta } \right\} \right].\end{eqnarray}
\end{widetext}
Note that the first term  is the Hawking temperature of the BTZ
black hole, while the second term is the $\theta$ correction.

The correction term of  surface gravity is given by
\begin{widetext}
\begin{eqnarray}\kappa \approx \frac{\sqrt{M}}{  l } - \left[ \frac{M \sqrt{M} l}{2 \theta } exp \left\{-\frac{M l^2}{4
\theta } \right\} + \frac{ \sqrt{M}  }{   l  } exp \left\{-\frac{M
l^2}{4 \theta } \right\} \right]\end{eqnarray}
\end{widetext}

\section{The Geodesics }\noindent
From the  geodesics equation
\begin{equation}
               \frac{d^2 x^\mu}{d\tau^2} + \Gamma^\mu_{\nu\lambda}
               \frac{d x^\nu}{d\tau}\frac{d x^\lambda}{d\tau}=0,
               \end{equation}
we obtain \cite{FDC}
\begin{equation}\label{E:geo1}
               \frac{1}{f(r)}\left(\frac{d r}{d\tau}\right)^2
               = \frac{E^2}{f(r)} - \frac{p^2}{r^2} +  L,
               \end{equation}
\begin{equation}\label{E:geo2}
               r^2\left(\frac{d \phi}{d\tau}\right) =  p,
               \end{equation}
and
\begin{equation}\label{E:geo3}
                \frac{d t}{d\tau} = \frac{E}{f(r)},
               \end{equation}
where   $ f(r) = -M+ 2M~\text{exp}
\left(-\frac{r^2}{4\theta}\right) - \Lambda r^2$  and the constants
$E$ and $p$ are identified as the energy per unit mass and angular
momentum, respectively. Here $\tau$ is the affine parameter and $L$
is the Lagrangian having values 0 and $-1$, respectively, for
massless and massive particles.

From the geodesic Eq. (\ref{E:geo1}) we can write
\begin{equation}\label{E:geo4}
\frac{1}{2}\left(\frac{dr}{d\tau}\right)^2 =
\frac{1}{2}\left[E^2+f(r)\left(L-\frac{p^2}{r^2}\right)\right].
\end{equation}

Now, comparing Eq. (\ref{E:geo4}) with $\frac{\dot{r}^2}{2} +
V_{eff} = 0 $, the effective potential can be written
\begin{equation}\label{E:V1}
V_{eff} =
-\frac{1}{2}\left[E^2+f(r)\left(L-\frac{p^2}{r^2}\right)\right].
\end{equation}

\subsection{Null geodesics }\noindent
For  massless particles, i.e., for photons, we have
$L=0$, and the corresponding effective potential is
\begin{equation} V_{eff} =
-\frac{E^2}{2}+\frac{p^2}{r^2} \left[-M+ 2M~\text{exp}
\left(-\frac{r^2}{4\theta}\right) - \Lambda r^2\right].
\end{equation}
As $ r \rightarrow 0 $, the effective potential
$V_{eff} (r)$ becomes very large, but it  approaches
$-\frac{E^2}{2}- \Lambda p^2$ as $ r \rightarrow \infty $.
At the horizons, $V_{eff} = - \frac{E^2}{2}$.

The shape of  $V_{eff} (r)$, shown in Fig. 7, indicates
that a photon will fall into a black hole \cite{FDC}. Taking
various values for the masses does not alter the nature of
the geodesics.

\begin{figure}
\includegraphics[scale=.3]{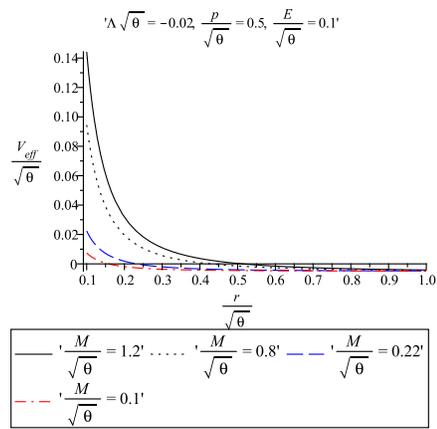}
\caption{Plot for $V_{eff} \sqrt{\theta}$ vs.
$\frac{r}{\sqrt{\theta}}$ for different values of
$\frac{M}{\sqrt{\theta}}$. Representation: the solid curve for $M=1.2
\sqrt{\theta}$,  dotted curve for $M=0.8 \sqrt{\theta}$,  dashed
curve for $M=0.214 \sqrt{\theta}$,  and chain curve for $M=0.1
\sqrt{\theta}$. The shape of the $V_{eff} \sqrt{\theta}$ indicates
that a photon will fall into black hole.}
\end{figure}

\subsection{Time-like geodesics }\noindent
For  massive particles, $L=-1$, and  the corresponding
effective potential is
\begin{multline}\label{E:V2}
V_{eff} =
-\frac{E^2}{2}\\+\left(1+\frac{p^2}{r^2}\right) \left[-M+
2M~\text{exp} \left(-\frac{r^2}{4\theta}\right) - \Lambda
r^2\right],
\end{multline}
shown in Fig. 8.  The effective potential becomes very large as
$r\rightarrow 0$, as well as when  $ r \rightarrow \infty $. At
the minimal horizon $r_0$, it assumes the constant value $V_{eff}
= -\frac{E^2}{2}$, while Fig. 9 shows that the roots of the
$V_{eff}$ coincide with the horizons. Also, the shape of the
effective potential indicates that the particle can move only
inside the black hole. Since the effective potential assumes
negative values between the horizons, the particle is confined to
the region between the two horizons, and, as a result, cannot hit
the singularity. Finally, observe that the minimum of $V_{eff}
(r)$ occurs between the horizons, so that  stable circular orbits
are going to exist.

\begin{figure}
\includegraphics[scale=.3]{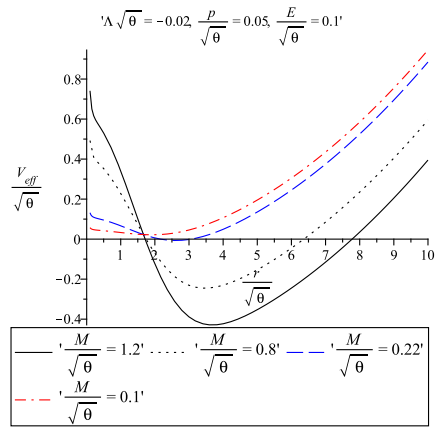}
\caption{Plot for $V_{eff} \sqrt{\theta}$ vs.
$\frac{r}{\sqrt{\theta}}$ for different values of
$\frac{M}{\sqrt{\theta}}$ . Representation: the solid curve for $M=1.2
\sqrt{\theta}$,  dotted curve for $M=0.8 \sqrt{\theta}$,  dashed
curve for $M=0.214 \sqrt{\theta}$,  and chain curve for $M=0.1
\sqrt{\theta}$.  The effective potential has a minimum between two
horizons, i.e., stable circular orbits do exist. }
\end{figure}

\begin{figure}
\includegraphics[scale=.3]{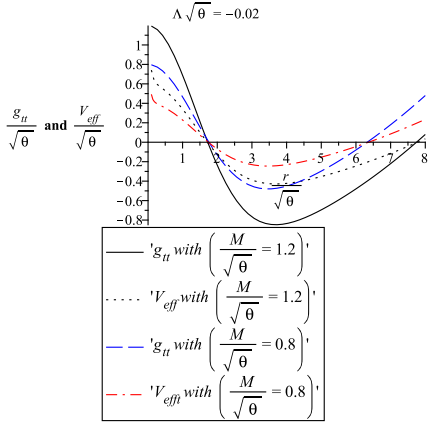}
\caption{Plot for $V_{eff} \sqrt{\theta}$ vs.
$\frac{r}{\sqrt{\theta}}$ for different values of
$\frac{M}{\sqrt{\theta}}$. Representation: the solid curve for $M=0.8
\sqrt{\theta}$,  dashed curve for $M=0.214 \sqrt{\theta}$, and chain
curve for $M=0.1 \sqrt{\theta}$.  The effective potential has a
minimum between two horizons, i.e., stable circular
orbits do exist. }
\end{figure}

\section{Test particles }\noindent
Let us consider a test particle having mass $m_0$  and moving in
the gravitational field of the BTZ black hole inspired by
noncommutative geometry and described by the metric ansatz
(\ref{E:line1}). The Hamilton-Jacobi [H-J] equation for the test
particle is \cite{fr1,fr2}

\begin{equation}
-\frac{1}{f} \left( \frac{\partial S}{\partial t}\right)^2  + f
\left( \frac{\partial S}{\partial r} \right)^2 +
\frac{1}{r^2}\left( \frac{\partial S}{\partial \phi}\right)^2 +
m_0^2 =0.
  \end{equation}
As there is no explicit dependence on $t$ and $\phi$, let us
choose the $H$-$J$ function $ S $ as \cite{fr1,fr2}
\[ S(t,r,\theta,\phi) = -E.t + S_1(r) + {p. \phi}, \]
where $E$  and $p$ are identified as the energy and
angular momentum of the particle.

The radial velocity of the particle is given by
\begin{equation}
 \frac{ dr}{ dt} = \frac{f^{\frac{3}{2}}}{E}
 \sqrt{\left[ \frac{E^2}{f} -m_0^2 -\frac{p^2}{r^2}
 \right]}.
\end{equation}
For detailed calculations, see Refs. \cite{fr1,fr2}.

The turning points of the trajectory are determined from
$\frac{dr}{dt} = 0 $ and, as a consequence, the potential
curve  is given by
\begin{equation} V(r) \equiv \frac{E}{m_0} =
 \sqrt{f} \left( 1+\frac{p^2}{m_0^2r^2} \right)^{1/2}.
 \end{equation}
The extremals of the potential curve are the solutions of the
equation $\frac{dV}{dr} = 0$
and are found to be
\begin{multline*}
\frac{dV}{dr} = - \frac{2 p^2}{m_0 r^3}\left[ -M+ 2M~\text{exp}
\left(-\frac{r^2}{4\theta}\right) - \Lambda r^2\right]
\\+ \left( 1+ \frac{ p^2}{m_0 r^2}\right) \left[
-\frac{Mr}{\theta} ~\text{exp} \left(-\frac{r^2}{4\theta}\right) -
2 \Lambda r\right] =0.
\end{multline*}
While difficult to tell from the equation, the plot of
$\frac{dV}{dr}$, given in Fig. 10, shows that real positive
solutions exist wherever $\frac{1}{\sqrt{\theta}}\frac{dV}{dr}$
cuts the $\frac{r}{\sqrt{\theta}}$-axis. Hence  bound orbits
for the test particles exist. In other words,  the test particles
can be trapped by BTZ black holes inspired by noncommutative
geometry.

\begin{figure}
\includegraphics[scale=.3]{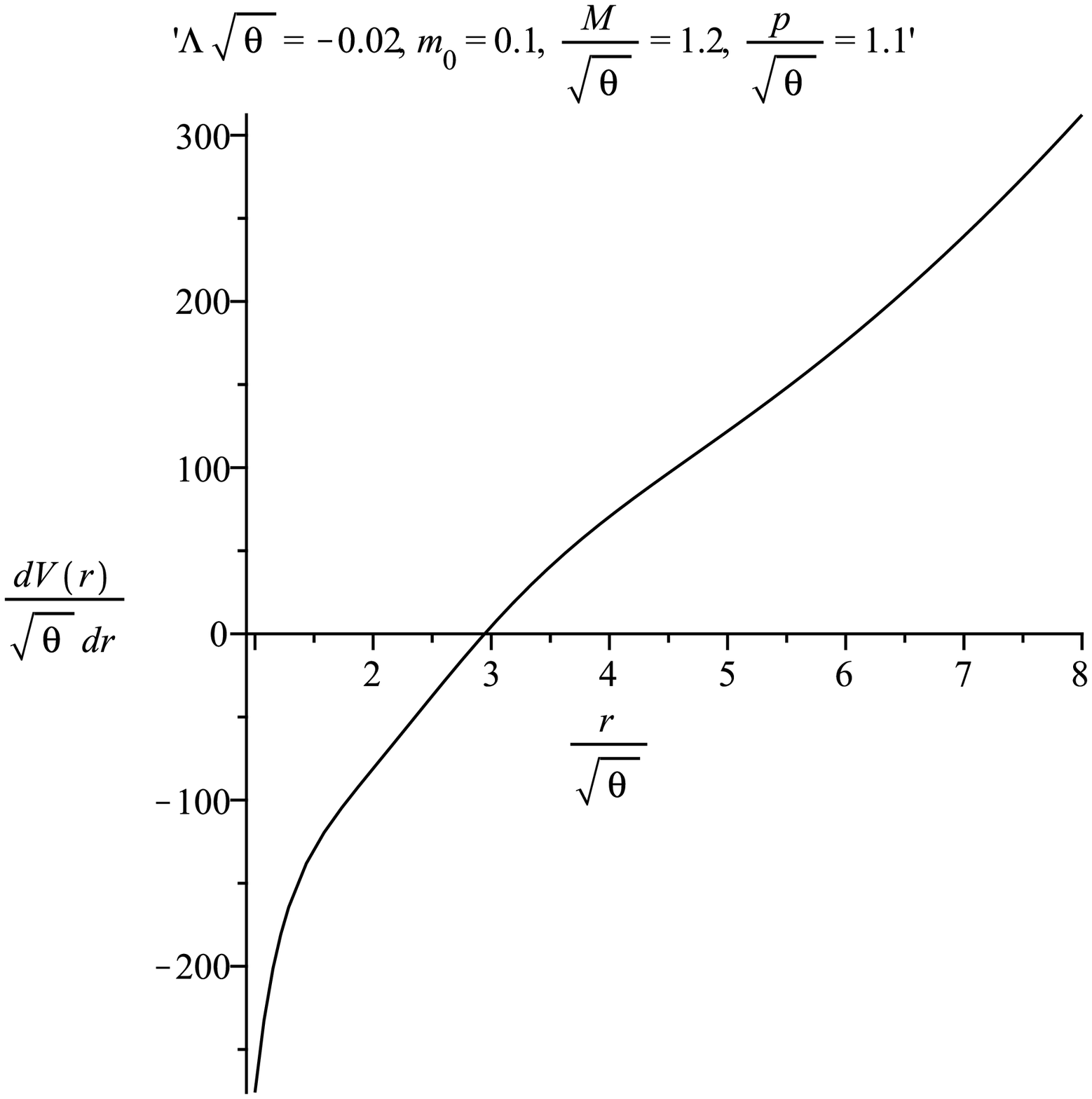}
\caption{Plot for $\frac{dV}{dr} \sqrt{\theta}$ vs.
$\frac{r}{\sqrt{\theta}}$. Note that  $\frac{dV}{dr}
\sqrt{\theta}$} is zero for certain value of
$\frac{r}{\sqrt{\theta}}$. This implies that $V(r)$ has at
least one extremal.
\end{figure}

\section{Conclusion}\noindent
This paper investigates the properties of a BTZ black hole
constructed from the exact solution of the Einstein field
equations in a $(2 + 1)$-dimensional anti-de Sitter spacetime in
the context of noncommutative geometry.

It was found that a BTZ black hole has two horizons, no horizons
or a single horizon $r=r_0$ corresponding to a minimal mass $M=
M_0$. In this connection we note the comments by Mazharimousavi et
al. \cite{Mazharimousavi}: ``It is well-known that unlike its
chargeless version the charged Banados-Teitelboim-Zanelli (BTZ)
black hole solution in (2 + 1)-dimensional spacetime is
singular''. Thus they construct a charged, regular extension of
the BTZ black hole solution by employing nonlinear Born-Infeld
electrodynamics, supplemented with the Hoffmann term and gluing
different spacetimes. However, our observation is that even the
noncommutative geometry inspired BTZ black hole is not free from
any singularity.

Beside this, in the present paper we continue our investigation
with a discussion of Hawking temperature, entropy and heat
capacity. We observe that the noncommutativity leads to the same
minimal radius $r_0$ at which the black hole cools down to
absolute zero. A discussion of the geodesic structure leads to the
effective potential for both massless and massive particles. It is
shown that photons will fall into the black hole, while massive
particles are trapped between the two horizons.  The use of the
Hamilton-Jacobi equation confirms that bound orbits are possible
for test particles.

\section*{Acknowledgments} FR and SR are thankful to the authority of
Inter-University Centre for Astronomy and Astrophysics, Pune,
India for providing them Visiting Associateship under which a part
of this work was carried out. FR is also thankful   UGC, Govt. of
India under research award scheme,  for providing financial
support. We are very grateful to an anonymous referee for his/her
insightful comments that have led to significant improvements,
particularly on the interpretational aspects.

\end{document}